\documentclass[final]{raa06}           
\usepackage{graphicx,times}             
\usepackage{natbib}
\usepackage{amssymb,amsmath}
\bibpunct{(}{)}{;}{a}{}{,}
\usepackage[colorlinks=true, citecolor=blue]{hyperref}%

\usepackage{graphicx,kantlipsum,setspace}
\usepackage{caption}
\usepackage{graphics,epsf}
\usepackage{amsmath}                
\usepackage{amsfonts}               
\usepackage{amssymb}                
\usepackage{epsfig}                 
\usepackage{appendix}
\usepackage{graphicx}
\usepackage{float}
\usepackage{color}
\usepackage{multirow}
\usepackage{colortbl}
\usepackage[para,online,flushleft]{threeparttable}
\usepackage{xcolor}

\hypersetup{citecolor=blue, 
            linkcolor=red, 
            menucolor=blue, 
            urlcolor=blue}  

\newcommand{\km}{{~\rm km}}
\newcommand{\s}{{~\rm s}}

\begin{document}

   \title{Jets are the most robust observable ingredient of common envelope evolution
}

   \volnopage{Vol.0 (20xx) No.0, 000--000}      
   \setcounter{page}{1}          

   \author{Noam Soker
    }

   \institute{Department of Physics, Technion, Haifa, 3200003, Israel;   {\it   soker@physics.technion.ac.il}\\
\vs\no
   {\small Received~~20xx month day; accepted~~20xx~~month day}}
\abstract{
I examine images of 50 planetary nebulae (PNe) with observable post-common envelope evolution (CEE) binary central stars and find that jets are about 40 percent more common than dense equatorial outflows. Because, in some cases, energetic jets can compress an equatorial outflow and because fast jets might disperse early in the PN evolution and avoid detection, the CEE process is likelier to launch jets than to eject a dense equatorial outflow by a larger factor than 1.4. In most cases, the companion, mainly a main sequence star, launches the jets as it accretes mass from the envelope of the giant star. By CEE jets, I also refer to jets launched shortly before the onset of the CEE, likely a grazing envelope evolution phase, and shortly after the CEE. The jets and the accretion of mass by the companion before, during, and after the CEE affect envelope mass removal and the final orbital separation. Most numerical simulations of the CEE ignore jets, and those that include jets omit other processes. Despite the considerable progress in the last decade with tens of hydrodynamical simulations of the CEE, we are still far from correctly simulating the CEE. Including jets in simulations of the CEE requires heavy computer resources, but it must be the next step.  
}

 \authorrunning{N. Soker}            
\titlerunning{Jets are the most robust observables of CEE}  
   
      \maketitle
\section{Introduction}
\label{sec:Introduction}

The opposite (to the center) pairs of lobes, clumps (also called ansae), bubbles, or `ears,' in many planetary nebulae (PNe), suggest morphological shaping by pairs of jets (e.g., \citealt{Morris1987, Soker1990AJ, SahaiTrauger1998, AkashiSoker2018,   EstrellaTrujilloetal2019, Tafoyaetal2019, Balicketal2020, RechyGarciaetal2020, Clairmontetal2022, Danehkar2022, MoragaBaezetal2023, Derlopaetal2024, Mirandaetal2024, Sahaietal2024}; for an alternative scenario see
\citealt{Baanetal2021}). Many of these studies consider a scenario where a main sequence companion accretes mass and launches the pairs of jets. The main sequence star accretes mass from the envelope of the asymptotic giant branch (AGB) progenitor of the PN. The same holds for rare cases of PNe that red giant branch (RGB) stars form (for PNe from RGBs, see e.g., \citealt{Hillwigetal2017, Sahaietal2017, Jonesetal2020, Jonesetal2022, Jonesetal2023}). 

There are more than a hundred bipolar and elliptical PNe, including some post-asymptotic giant branch nebulae and pre-PNe, having a central binary system (e.g., \citealt{Miszalski2019ic, Oroszetal2019, Jones2020Galax, Jones2025}). The short orbital periods of these binary systems show that they have experienced a common envelope evolution (CEE) phase. In post-AGB systems and PNe with long orbital periods, months to years { (e.g., \citealt{vanWinckeletal2014}), } the systems might have experienced the grazing envelope evolution (GEE). The morphologies of many of these suggest that jets shaped these PNe, pointing at the direct link between CEE and jet-launching. In this study, I quantify this occurrence rate and compare it to the occurrence rate of equatorial outflows (ring/torus/disk). 

During the CEE of AGB and RGB stars, these cool giants' high mass loss rate implies heavy dust formation, obscuring the system.
Dust's high opacity might facilitate envelope removal (e.g., \citealt{Soker1998dust, Soker2000dust, Luetal2013, GlanzPerets2018, Iaconietal2019, Iaconietal2020, Reichardtetal2020, BermudezBustamanteetal2024a}). 
\cite{BermudezBustamanteetal2024b} find that in the case of a relatively massive companion in the CEE, most dust forms in the unbound ejecta and hence has little effect on mass ejection.
Because the CEE cannot be observed directly, many studies have aimed at determining the properties of the CEE phase by examining post-CEE binary systems.
Some studies (e.g., \citealt{IaconiDeMarco2019} and references herein) estimate the efficiency of mass removal by deposition of orbital energy, ignoring other energy sources (the $\alpha_{\rm Ce}$ parameter). 
The larger radii of some main sequence companions at the center of planetary nebulae with post-CEE binary systems suggest that the companion accreted mass during the CEE (e.g., \citealt{Jonesetal2015}).  \cite{MichaelyPerets2019} and \cite{Igoshevetal2020} constrain the duration of CEE from a tertiary star orbiting some post-CEE binaries. 

In this study, I limit myself to the simple question of the fraction of post-CEE binaries with jet signatures. As jets can have signatures on PN morphologies, I examine PNe with central binary stars (Section \ref{sec:Jets}). In Section \ref{sec:Implications} I discuss some implications and relations to the CEE. 
I summarize this short study with a firm conclusion in Section \ref{sec:Summary}. 

\section{The sample of post-CEE PNe with jets}
\label{sec:Jets}

I examined the catalog of PNe with a binary central star built and maintained by David Jones\footnote{\url{https://www.drdjones.net/bCSPN/}}
\citep{JonesBoffin2017, BoffinJones2019, Jones2025}. 
I do not consider the uncertain cases in that catalog (marked by * or ** there). 
This leaves 108 PNe. I could find high-quality images to decide whether jets exist in 50 PNe. Out of these 50 PNe, I saw clear indications for jets or morphologies robustly suggesting jets in 28 PNe, marked by the number `1' in the third column of Table \ref{Tab:Table1}. In 8 PNe, I found morphologies that likely suggest jets, marked by `0.5' in the third column. The fourth column of Table \ref{Tab:Table1} indicates the presence (by `1') or not (by `0') of an equatorial outflow or a ring/torus; a number `0.5' indicates likely equatorial mass concentration.
A subscript `j' near the reference indicates that this reference suggests jets (with possible references to earlier claims for jets). An up-script points to another reference that claims for jets. Otherwise, the claim for jets is in this study.  { The subscript `r' similarly refers to the evidence for a ring (equatorial outflow).  }  
\begin{table*}
  \caption{Planetary nebulae with binaries and jets}
\begin{minipage}{0.5\textwidth}  
    \begin{tabular}{| p{1.6cm} | p{2.2cm} | p{1.6cm}| p{0.4cm}|  }
\hline  
\textbf{{PN}}  & \textbf{P (day)} & \textbf{Jets} & \textbf{R} \\
\hline  
NGC 1360 & 142 $^{\rm [Mi18n]}$    &  1 $^{\rm [GD08]_{\rm j}}$ &  0 \\   
MyCn 18  & 18.15 $^{\rm [Mi18m]}$  &  1 $^{\rm [Oc00]_{\rm j}}$ &  0 \\ 
NGC 2346 &  16.00 $^{\rm [Br19]_{\rm r}}$  & 0.5 $^{\rm [Go19]}$ & 1  \\
Sp 3     &  4.81 $^{\rm [Mi19s]}$  &  1 $^{\rm [Mi19s]_{\rm j,r}}$ & 0.5 \\
IC 4776  &  3.11 $^{\rm [Mi19i]}$ &  1 $^{\rm [So17]_{\rm j}}$  & 0.5    \\
NGC 7293 &  2.77 $^{\rm [Al20]}$ &    1 $^{\rm [Me13]_{\rm j}}$  &  0    \\
IC 2149  &  2.63877 $^{\rm [Al24]_{\rm r}}$  & 1 $^{\rm [Va02]^{\#1}}$ &  1 \\
IC 4593  &  2.50369 $^{\rm [Al21]}$  & 1 $^{\rm [To20]_{\rm j}}$  &  0   \\
NGC 2392 &  1.90 $^{\rm [Mi19n]}$  &  1 $^{\rm [Gu21]_{\rm j}}$  &  0   \\
Fg 1     &  1.195 $^{\rm [We18]}$  &  1 $^{\rm [Lo93]_{\rm j}}$  &  0   \\
Necklace &  1.16 $^{\rm [Mi13]_{\rm j}}$  &  1$^{\rm [Co11]_{\rm r}}$  & 1  \\
Hen 2-283& 1.15 $^{\rm [We18]}$  & 0.5 $^{\rm [We18]}$  & 0.5    \\
Hen 2-161& $\simeq 1$ $^{\rm [Jo15]_{\rm j}}$ & 1$^{\rm [Sa11]_{\rm r}}$& 1 \\
Abell 65 &  1 $^{\rm [Hi15]}$  & 0.5 $^{\rm [Hu13]_{\rm j}}$  & 0    \\
HaWe 8   & 0.99 $^{\rm [Bh24]}$  &  0.5 $^{\rm [Zu24]}$  & 0    \\
K 1-2 & 0.676 $^{\rm [Ex03]}$  & 1 $^{\rm [Co99]_{\rm j}}$  &  0   \\
M 2-19 & 0.670 $^{\rm [Mi08]}$  & 0.5 $^{\rm [Mi08]_{\rm r}}$  & 1    \\
Hen 2-11 & 0.609 $^{\rm [Jo14]}$  & 1 $^{\rm [Jo14]_{\rm j}}$  & 1    \\
\hline  
\end{tabular}
\end{minipage} \hfill
\begin{minipage}{0.5\textwidth}
    \begin{tabular}{| p{1.6cm} | p{2.2cm} | p{1.6cm}| p{0.4cm}|  }
\hline  
\textbf{{PN}}  & \textbf{P (day)} & \textbf{Jets} &\textbf{R} \\
\hline  
M 3-16 & 0.574 $^{\rm [Mi08]}$  & 1 $^{\rm [Mi08]_{\rm j,r}}$  &  0.5    \\
ETHOS 1 & 0.53 $^{\rm [Mi11e]}$  & 1 $^{\rm [Mi11e]_{\rm j}}$  & 0.5 \\
Hen 2-84 & 0.485645 $^{\rm [Hi22]}$  & 1 $^{\rm [Hi22]_{\rm r}}$  & 1 \\
Abell 46 & 0.472 $^{\rm [Co15]}$  & 0.5 $^{\rm [Co15]}$  & 0 \\
Abell 63 & 0.465 $^{\rm [Co15]}$  & 1 $^{\rm [Mt07]_{\rm j}}$  & 0 \\
NGC 6326 & 0.37 $^{\rm [We18]}$  & 1 $^{\rm [Mi11j]_{\rm j}}$  & 0.5 \\
Ou 5 & 0.364 $^{\rm [Co15]}$  & 1 $^{\rm [Co14]_{\rm j,r}}$  &  0.5 \\
M 2-46 & 0.3192 $^{\rm [Bh24]}$  & 1 $^{\rm [Ma96]^{\#2}}$  &  1   \\
Sab 41 & 0.297 $^{\rm [Mi09]}$  & 1 $^{\rm [Mi11a]_{\rm j,r}}$  &  1   \\
Bl 3-15 & 0.270 $^{\rm [Mi09]}$  & 1 $^{\rm [Mi09]_{\rm j}}$  &  0   \\
PN G283 & 0.25 $^{\rm [Jo20]}$  & 1 $^{\rm [Jo20]}$  &  0 \\
H 2-29 & 0.244 $^{\rm [Mi09]}$  & 0.5 $^{\rm [Mi09]_{\rm j,r}}$  &  1 \\  
Abell 41 & 0.226 $^{\rm [Jo10]}$  & 1 $^{\rm [Jo10]_{\rm r}}$  &  1 \\
Hen 2-428 & 0.176 $^{\rm [SG15]}$  & 1 $^{\rm [SG15]_{\rm r}}$  &  1 \\
NGC 6778 & 0.15 $^{\rm [Jo16]}$  & 1 $^{\rm [GM12]_{\rm j}^{\#3}}$  &  0.5 \\
Hen 2-155 &0.148 $^{\rm [Jo15]}$  & 1 $^{\rm [Jo15]_{\rm j}}$  &  0  \\
Pe 1-9 & 0.140 $^{\rm [We18]}$  & 0.5 $^{\rm [Sc92]}$  &  0   \\
M 3-1 & 0.127 $^{\rm [jo19]}$  & 1 $^{\rm [Jo19]_{\rm j}}$  &  0   \\
\hline  
     \end{tabular}
     \end{minipage} \hfill
  \label{Tab:Table1}\\
\small 
Note:  36 PNe with binary central stars and high-quality images indicating jets. The first column is the name of the PN, and the second is the orbital period of the central star and the source for it, as appears in the site built and maintained by David Jones (e.g., \citealt{Jones2025}). The third column indicates robust morphological indications for jets (marked by `1') or likely morphological indications for jets (`0.5'), with the source of the image I used. A subscript `j' in one of the two references in a row indicates that the binary or the image reference claims for jets, while an up-script mentions another reference to support jets. Otherwise, the claim for jets is in this study. { The same holds for a subscript `r' for the indication of an equatorial dense gas (ring). } The fourth column indicates robust (`1'), likely (`0.5'), or no (`0') morphological indications for an equatorial outflow, i.e., disk, torus, or ring.  
\newline
Abbreviation: P: orbital period; PN: planetary nebula; R: ring (including torus or a disk); PN G 283: PN G283.7-05.1.
\newline
References: 
Al24:    \cite{Alleretal2024};
Al20:    \cite{Alleretal2020};
Bh24;   \cite{Bhattacharjeeetal2024};
Br19:    \cite{Brownetal2019};
Co99:   \cite{Corradietal1999}:
Co11:   \cite{Corradietal2011};
Co14:   \cite{Corradietal2014};
Co15:   \cite{Corradietal2015};
Ex03:   \cite{Exteretal2003};
GD08:   \cite{GarciaDiazetal2008};
Go19:    \cite{GomezMunozetal2019};
Gu21:    \cite{Guerreroetal2021};
GM12:   \cite{GuerreroMiranda2012};
Hi15: \cite{Hillwigetal2015};
Hi22:  \cite{Hillwigetal2022H};
Hu13:   \cite{Huckvaleetal2013};
Jo10:   \cite{Jonesetal2010}; 
Jo14:    \cite{Jonesetal2014Hen};
Jo15:   \cite{Jonesetal2015};
Jo16:    \cite{Jonesetal2016};
Jo19:   \cite{Jonesetal2019};
Jo20:   \cite{Jonesetal2020};
Lo93:    \cite{Lopezetal1993};
Ma96:   \cite{Manchadoetal1996};
Me13:   \cite{Meaburnetal2013};
Mt07:  \cite{Mitchelletal2007};
Mi08:   \cite{Miszalskietal2008};
Mi09:   \cite{Miszalskietal2009II};
Mi11e:   \cite{Miszalskietal2011E};
Mi11j:   \cite{Miszalskietal2011AA};
Mi11a:  \cite{Miszalskietal2011apn5};
Mi13:   \cite{Miszalskietal2013};
Mi18n:  \cite{Miszalskietal2018ngc1360};
Mi18m: \cite{Miszalski2018mycn18};
Mi19s:  \cite{Miszalskietal2019sp};
Mi19i:  \cite{Miszalski2019ic};
Mi19n: \cite{Miszalskietal2019ngc2392};
Oc00:    \cite{OConnoretal2000};
Sa11:   \cite{Sahaietal2011};
Sc92:   \cite{Schwarzetal199};
SG15:   \cite{SantanderGarciaetal2015};
So17:    \cite{Sowickaetal2017};
To20:   \cite{Toalaetal2020};
Va02:    \cite{Vazquezetal2002};
We18:   \cite{Wessonetal2018};  
Zu24: \url{https://brentenriegel.at/ergebnisse/a240103-btb-hawe-8-planetarischer-nebel};
{ \#1: claim for jets in \cite{Balicketal1993}. \#2: claim for a ring in \cite{Manchadoetal1996b}. \#3: claim for a ring in \cite{Miszalskietal2011AA}. }
%
 %
\end{table*}

\begin{figure*} 
\begin{center}
\includegraphics[trim=0cm 10.0cm 0.0cm 0cm,scale=0.70]{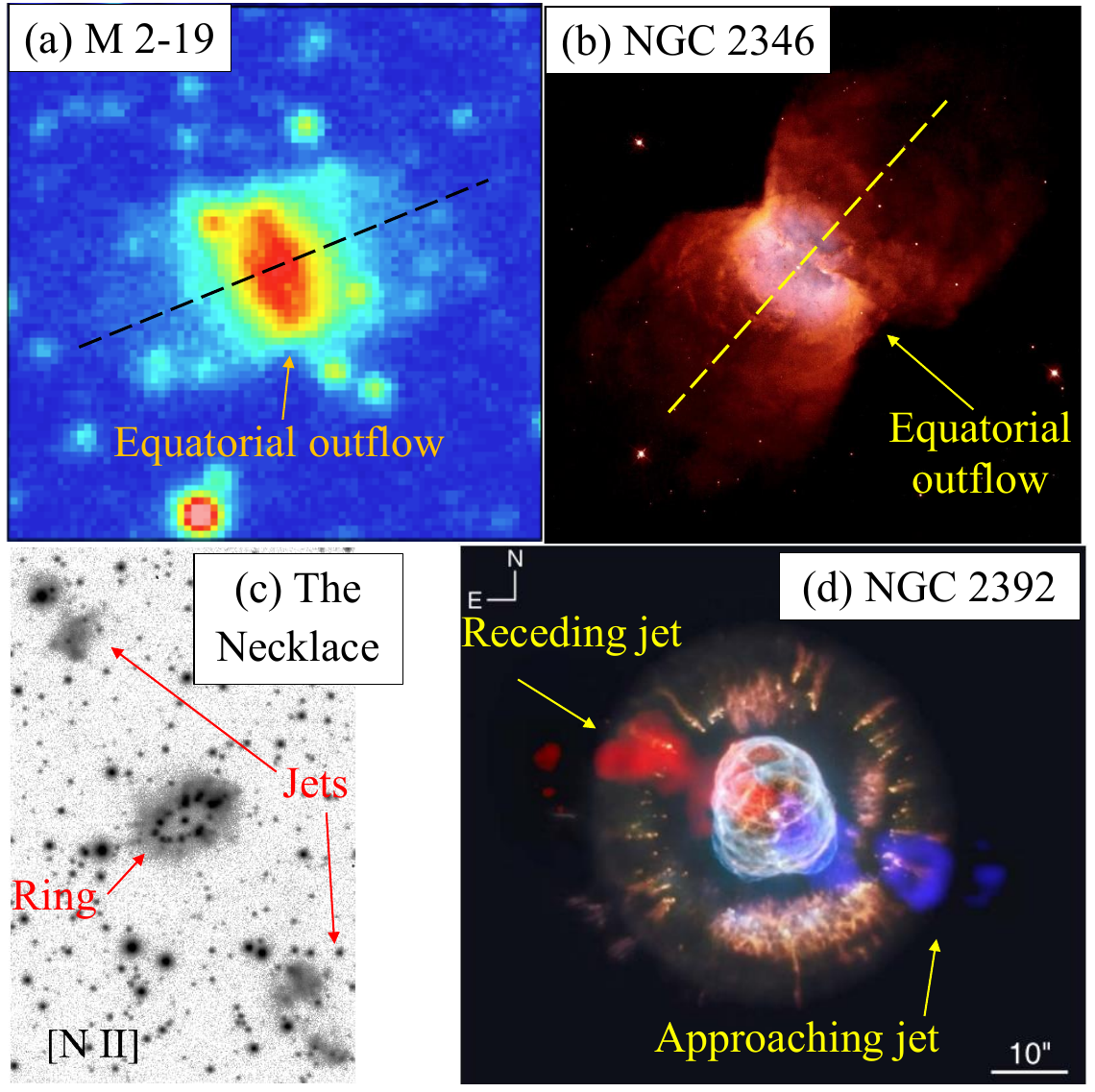} 
\caption{Four PNe demonstrating jets and equatorial outflows. 
(a) An image of M 2-19 adapted from \cite{Miszalskietal2008}. (b) An HST image of NGC 2346. In panels (a) and (b) I marked the equatorial outflow and added the dashed line in each panel to indicate the polar axis. These two PNe demonstrate equatorial mass concentration, hence the value `1' in the fourth column of Table \ref{Tab:Table1}.There are no direct indications of jets, only indirect indications by the openings to the two polar directions, hence the number `0.5' in the third column. 
(c) An image of The Necklace adapted from \cite{Corradietal2011} clearly demonstrates the existence of jets and an equatorial ring. 
(d) An image of NGC 2392 from \cite{Guerreroetal2021} demonstrating that, in some cases, a deep analysis is required to reveal the jets. The blue regions are the approaching jets at $-190 \km \s^{-1}$ to $-150 \km \s^{-1}$, while the red regions are the receding jet $+150 \km \s^{-1}$ to $+190 \km \s^{-1}$. 
The white regions near the center are the nebula at the velocity interval of $-135 \km \s^{-1}$ to $+135 \km \s^{-1}$. There is no indication of a dense equatorial outflow. }
\label{Fig:PNe}
\end{center}
\end{figure*}

Specifically, there are cases of directly observed jets, e.g., ETHOS 1 and the Necklace, while in some cases that have no observable jets, the bipolar structure robustly indicates jets, like the couple of misaligned lobes of M 2-46 and the extended lobes of Ou 5.  In some cases, the bipolar structure might be attributed to the collimation of fast wind by an equatorial dense gas; in those cases, I marked in the third column of Table \ref{Tab:Table1} the number 0.5 (although I think these are also shaped by jets).  
{ Figure \ref{Fig:PNe} presents four PNe that demonstrate some of the classifications. }

Out of the 36 PNe with jets, or likely jets, 12 have the robust structures of an equatorial outflow (disk/torus/ring), and 8 have a likely equatorial outflow.
One of the 14 PNe with high-quality images but without indications of jets has a robust indication of an equatorial outflow (HaTr 4), and four have likely indications (AMU 1; Sp 1; RWT 152; NGC 6337).
 Overall, 25 PNe out of the 50 have robust or likely indications for an equatorial outflow. I summarize these in Table \ref{Tab:Table2}, and discuss the implications in Section \ref{sec:Implications}. 
\begin{table}
  \caption{Distribution by percentage}
\begin{minipage}{0.5\textwidth}  
    \begin{tabular}{| p{2.0cm} | p{1.0cm} | p{1.0cm}| p{1.0cm}|  }
\hline  
\textbf{{}}  & {Robust } & likely & No \\
\textbf{{}}  & {jets} & Jets & jets \\
\hline  
Robust ring & 18 & 6 &  2  \\   
Likely ring & 14 & 2 &  8  \\
No ring     & 24 & 8 & 18  \\ 
\hline  
     \end{tabular}
     \end{minipage} \hfill
  \label{Tab:Table2}\\
\small 
Note:  Morphological classification by percentage of PNe with central binary stars and high-quality images to decide on the presence or not of jets; in total, 50 PNe (Table \ref{Tab:Table1} and text). The word `ring' indicates equatorial mass concentration, including a torus or a disk. 
%
 %
\end{table}

To increase this sample, I highly encourage further and more profound studies of PNe, for which I could not find images that would allow me to decide whether or not morphological indications for jets exist. 

\section{Implications for CEE}
\label{sec:Implications}
\subsection{Jetted-to-ring post-CEE PNe ratio}
\label{subsec:jets}

I base the statistical summary of Table \ref{Tab:Table2} on 50 PNe with binary systems and images that allow the determination of the existence or not of jets. While $72\%$ percent of PNe show robust and likely morphological indications for jets, only $50\%$ have morphological indications for a dense equatorial outflow in the form of disk/torus/ring. The directly observed ratio of jetted-to-ring ratio of post-CEE PNe is 
\begin{equation}
  Q_{\rm JR,obs} \simeq 1.4. 
    \label{eq:ObsRatio}
\end{equation}

However, I expect the jetted-to-ring ratio due to the CEE interaction itself to be much larger for two reasons. (1) { Detection biases. (1.1) } Equatorial outflows from CEE are dense and slow and stay bright even as they expand. On the other hand, jets expand fast and disperse on a shorter time scale. 
{ One example is the Necklace (panel c of Figure \ref{Fig:PNe}). The jets are fainter and at a much larger distance from the center than the ring. A similar PN but much older, by thousands of years, might reveal only its ring and not the jets. (1.2) A ring is bright and prominent from all directions. Jets are hard to detect when pointing at the observer. The PN NGC 2392 (panel c of Figure \ref{Fig:PNe}) is an example. Only a deep analysis with Doppler shift reveals its jets. }
 (2) In some PNe, I expect jets to shape an equatorial dense outflow rather than the direct CEE mass ejection. \cite{Akashietal2015} and \cite{Shiber2018} demonstrated the formation of such rings with three-dimensional hydrodynamical simulations. \cite{Akashietal2015} present the equatorial ring of SN 1987a and the Necklace as specific examples.
I marked the Necklace as having robust signatures of jets and a ring. However, the claim of \cite{Akashietal2015} is that the jets shaped the ring rather than direct CEE mass ejection. 
For the above reasons, I expect that jets have shaped a significant fraction of PNe that currently lack jet morphological signatures. { A quantitative study is for the future, with deeper analysis of more PNe. From the large number of PNe with barely observed jets (like the Necklace and NFC 2392), } I crudely estimate that the jetted-to-ring ratio due to the CEE mass ejection { is approximately twice as given from the inspection of images. } 
\begin{equation}
  Q_{\rm JR,CEE} \approx 2 \times Q_{\rm JR,obs} \gtrsim 3. 
    \label{eq:CEERatio}
\end{equation}
The determination of this ratio is for the future, as well as the inclusion of other processes that influence jets and equatorial mass outflows, like magnetic fields (e.g., \citealt{Nordhausetal2007}). 
{ I note that the possibility of missing an equatorial ring because it is not resolved is unlikely. Such a ring (torus) is very bright because it is dense and close to the center. The bright waist of Hen 2-84 (e.g., \citealt{Hillwigetal2022H}) is an example. }
At this time, I conclude that jets are more common than dense equatorial outflows concerning the CEE process. 

\subsection{The jets' source}
\label{subsec:source}

Precessing jets, such as in FG 1 (e.g., \citealt{Boffinetal2012})
and IC 4776 (e.g., \citealt{Sowickaetal2017}), 
and multipolar PNe, such as the two misaligned pairs of lobes, one inside the other in M 2-46, suggest that the symmetry axis of the jets can change over the formation time of the PNe. This implies that the source of the jets is not the common envelope (as simulated by, e.g., \citealt{GarciaSeguraetal2020, GarciaSeguraetal2021}) because the angular momentum of the common envelope does not change during the CEE of binary stars. An accretion disk around one of the two stars, the main sequence companion, or the core, launches the jets. Because in some cases, jets are pre-CEE (e.g., \citealt{Tocknelletal2014}), the typical case is that the main sequence companion launches the jets as it accreted mass from the giant's envelope before, during, or after the CEE phase.

Some hydrodynamic simulations of the CEE obtain a funnel in the common envelope along the angular momentum axis of the binary system, a funnel that might launch jets (e.g., \citealt{Chamandyetal2018, Chamandyetal2020, Zouetal2020, Zouetal2022, Morenoetal2022, Ondratscheketal2022, Vetteretal2024, GagnierPejcha2025}). Based on analytical calculations, the launch of jets from the common envelope funnel was suggested many years ago  \citep{Soker1992funnel}. However, the common envelope is not expected to precess and hence cannot form a point-symmetrical PN. Jets can change directions due to instabilities, but the two opposite jets change direction uncorrelated, hence do not form a point-symmetrical morphology (e.g., \citealt{Ondratscheketal2022}).  Moreover, the jets that, e.g., \cite{Ondratscheketal2022} and \cite{GagnierPejcha2025}, find in their simulations are too weak to explain the shaping of PNe, i.e., they bent close to the envelope. 

\cite{BlackmanLucchini2014} who examined the kinetic properties of axial-symmetric outflows from 19 pre-PNe, which were observed and analyzed by  \cite{Bujarrabaletal2001} and \cite{Sahaietal2008}, concluded that main sequence companions to the AGB progenitors of the pre-PNe should accrete at a very high rate to explain these kinetic properties. Accretion might in some pre-PNe be via a Roche lobe overflow (RLOF), or, more likely, for most or all pre-PNe, during a CEE and sometimes during a GEE. 
There are indications that the main sequence companion accretes mass during the CEE and/or shortly before the CEE, namely, during the GEE. 
An example is the larger radius than that of a main sequence star with an equal mass of the companion in Hen 2-155, which probably results from rapid mass accretion \citep{Jonesetal2015}.  

The relation of the shaping jets to the binary system is evident from the general alignment of the angular momentum axes of the central binaries of PNe with the PNe main symmetry axes \citep{Hillwigetal2016}.

\subsection{The relation of the jets to CEE}
\label{subsec:ratio}

Jets can be older, coeval, or younger than the main PN shell (e.g.,  \citealt{Tocknelletal2014, Guerreroetal2020, Kimeswengeretal2021}). 
In \cite{Soker2020Galax}, I summarized the different phases of jet launching, including the GEE, and presented supporting arguments for the GEE. I will not repeat that discussion, as the present paper deals only with observational signatures of jets. In general, when arguing for jets in the CEE, I refer also to jets that the companion launches during the GEE phase in cases where the GEE exists. I here discuss the energy budget that jets and the jet-launching companion introduce. 

Motivated by the present finding that jet-launching is the most robust observable of the CEE and cannot be ignored, I consider the possible implications to the energy budget.
The launching of jets requires the main sequence companion to accrete mass. The jets allow this accretion (e.g., \citealt{Shiberetal2016}) operating in a valve mechanism that releases pressure \citep{Chamandyetal2018}. Applying the virial theorem, the accretion process onto the main sequence companion releases energy of 
\begin{equation}
E_{\rm acc}=\frac{GM_{\rm MS} M_{\rm acc}} {2 R_{\rm MS}} ,
    \label{eq:AccretionE}
\end{equation}
where $M_{\rm MS}$ and $R_{\rm MS}$ are the companion's initial mass and average radius during the accretion process, and $M_{\rm acc}$ is the accreted mass. 
{ Recent studies suggest that main sequence stars can accrete mass at high rates with only minor expansion if jets remove the outskirts of the envelope (\citealt{BearSoker2025}; \citealt{Scolnicetal2025}).  }
The released orbital energy, if the leftover envelope mass is low, when the orbital radius is $a$, is 
 \begin{equation}
E_{\rm orb} (a)=\frac{GM_{\rm MS} M_{\rm core}} {2 a} ,
    \label{eq:OrbitE}
\end{equation}
where $M_{\rm core}$ is the giant's core mass. 
Accretion energy dominates over orbital energy in radii of 
\begin{equation}
a \gtrsim \frac{M_{\rm core}} {M_{\rm acc}} R_{\rm MS}= 10 \left( \frac{M_{\rm core}} { 20 M_{\rm acc}} \right) \left( \frac{R_{\rm MS}}{0.5 R_\odot} \right) R_\odot. 
    \label{eq:OrbitA}
\end{equation}
Typical values can be $M_{\rm MS}=0.15-0.4 M_\odot$,  $R_{\rm MS}=0.4-0.7 R_\odot$, and $M_{\rm core}=0.5-0.7 M_\odot$ { (e.g., \citealt{Hillwigetal2010, DeMarcoetal2011, Jonesetal2015}). I scale with a value of } 
$M_{\rm acc} \approx 0.03 M_\odot$. Jets might carry a fraction of 0.1-0.2 of the accretion energy. The rest is radiated away. Not all the energy deposited by the different processes (accretion, orbital, recombination) ends in unbinding the envelope; the convection can efficiently carry energy out where it is radiated away (e.g.,  \citealt{Sabachetal2017, WilsonNordhaus2019, WilsonNordhaus2020, WilsonNordhaus2022, Noughanietal2025}). I do not discuss recombination energy here (for a recent paper and references, see \citealt{Chamandyetal2024}; recombination seems to play a small role in envelope ejection). 

As it emerges from the envelope near the core, the main sequence companion might have a radius larger than its zero-age main sequence radius and is exposed to the core's radiation. It might lose some of the mass it accreted into the binary vicinity by expanding and filling its Roche-lobe and adding mass to a possible circumbinary disk. In any case, the ejection of this mass requires the main sequence companion to spiral in further, reducing the orbit by a fraction of $\simeq M_{\rm acc}/M_{\rm MS}$. Namely, the mass accreted before and during the CEE helped mass removal, which now absorbs energy and reduces the orbit.  

My point is that jets not only deposit energy and shape planetary nebulae, but they also imply a significant mass accretion by the main sequence companion. This accretion process might facilitate mass removal and, at the post-CEE, some further spiraling in. Finding observational indications for mass accretion is more challenging than launching jets. But the jets imply the mass accretion process.

\section{Summary}
\label{sec:Summary}

I examined 50 PNe with observable post-CEE central binary systems and high-quality images in this short study. I found that jets are more common than dense equatorial outflow in the ejecta of post-CEE binaries (Table \ref{Tab:Table2}), by about $40\%$ (equation \ref{eq:ObsRatio}). However, because jets might disperse and become non-detectable and because an equatorial dense outflow might result from the jets rather than a CEE equatorial ejection (Section \ref{subsec:jets}), I estimated that from the CEE process itself, this ratio is larger even (equation \ref{eq:CEERatio}).

The robustness of jets' association with CEE mass ejection implies that numerical codes must incorporate jets' roles in CEE simulations. I do not claim that, in all cases, the companion launches jets during the entire CEE phase. The companion might launch the jets outside the envelope, mainly in a GEE. In many cases, I expect the accretion disk with the jets it launches to survive into the CEE, at least in the outer zones. The companion can also launch jets in the post-CEE phase (see \citealt{Soker2020Galax} for the different phases of jet-launching). These imply that comparisons of CEE simulations to observations, like the process of mass ejection and the final orbital radius of the core-companion binary system, should consider the role of jets. 

 Most of the tens of CEE simulations (e.g., just from the last decade, \citealt{Nandezetal2014, Staffetal2016MN, Kuruwitaetal2016, Ohlmannetal2016a,  Iaconietal2017b, Chamandyetal2019, LawSmithetal2020, GlanzPerets2021a, GlanzPerets2021b, GonzalezBolivaretal2022, GonzalezBolivaretal2024, Lauetal2022a, Lauetal2022b,  BermudezBustamanteetal2024, Chamandyetal2024, GagnierPejcha2024,  Landrietal2024, RosselliCalderon2024, Vetteretal2024}) do not include the role of jets. Those simulations that do include jets cover a small fraction of the orbital period or omit other processes (e.g., \citealt{MorenoMendezetal2017, ShiberSoker2018, LopezCamaraetal2019, Shiberetal2019, LopezCamaraetal2020MN, Hilleletal2022, Hilleletal2023, LopezCamaraetal2022, Zouetal2022, Soker2022Rev, Schreieretal2023, Gurjareta2024eas, ShiberIaconi2024}).
I list many (but not all) of the last decade's simulations to emphasize that despite the considerable progress and many studies in the last decade, we are still far from correctly simulating the CEE. 

Including jets in CEE simulation requires heavy computational power, which not all research groups have (e.g., our group, whose recent proposal was rejected).

\section*{Acknowledgements}

 I thank Gunter Cibis and an anonymous referee for their helpful comments.  
I heavily used the catalog of PNe with a binary central star built and maintained by David Jones (see text) and the Planetary Nebula Image Catalogue (PNIC) of Bruce Balick (\citealt{Balick2006}; \url{https://faculty.washington.edu/balick/PNIC/}). 
A grant from the Pazy Foundation supported this research.

\label{lastpage}


\begin{thebibliography}{}
\expandafter\ifx\csname natexlab\endcsname\relax\def\natexlab#1{#1}\fi
\providecommand{\url}[1]{\href{#1}{#1}}
\providecommand{\dodoi}[1]{doi:~\href{http://doi.org/#1}{\nolinkurl{#1}}}
\providecommand{\doeprint}[1]{\href{http://ascl.net/#1}{\nolinkurl{http://ascl.net/#1}}}
\providecommand{\doarXiv}[1]{\href{https://arxiv.org/abs/#1}{\nolinkurl{https://arxiv.org/abs/#1}}}

\bibitem[{{Akashi} {et~al.}(2015){Akashi}, {Sabach}, {Yogev}, \& {Soker}}]{Akashietal2015}
{Akashi}, M., {Sabach}, E., {Yogev}, O., \& {Soker}, N. 2015, \mnras, 453, 2115, \dodoi{10.1093/mnras/stv1666}

\bibitem[{{Akashi} \& {Soker}(2018)}]{AkashiSoker2018} {Akashi}, M., \& {Soker}, N. 2018, \mnras, 481, 2754, \dodoi{10.1093/mnras/sty2479}

\bibitem[{{Aller} {et~al.}(2024){Aller}, {Lillo-Box}, \& {Jones}}]{Alleretal2024}
{Aller}, A., {Lillo-Box}, J., \& {Jones}, D. 2024, \aap, 690, A190, \dodoi{10.1051/0004-6361/202450942}

\bibitem[{{Aller} {et~al.}(2020){Aller}, {Lillo-Box}, {Jones}, {Miranda}, \& {Barcel{\'o} Forteza}}]{Alleretal2020} {Aller}, A., {Lillo-Box}, J., {Jones}, D., {Miranda}, L.~F., \& {Barcel{\'o} Forteza}, S. 2020, \aap, 635, A128, \dodoi{10.1051/0004-6361/201937118}

\bibitem[{{Baan} {et~al.}(2021){Baan}, {Imai}, \& {Orosz}}]{Baanetal2021}
{Baan}, W.~A., {Imai}, H., \& {Orosz}, G. 2021, Research in Astronomy and Astrophysics, 21, 275, \dodoi{10.1088/1674-4527/21/11/275}

\bibitem[{{Balick}(2006)}]{Balick2006}
{Balick}, B. 2006, {Planetary Nebula Image Catalogue: HST data}, HST Proposal ID 10933. Cycle 15

\bibitem[{{Balick} {et~al.}(2020){Balick}, {Frank}, \& {Liu}}]{Balicketal2020}
{Balick}, B., {Frank}, A., \& {Liu}, B. 2020, \apj, 889, 13, \dodoi{10.3847/1538-4357/ab5651}

\bibitem[\protect\citeauthoryear{Balick et al.}{1993}]{Balicketal1993} Balick B., Rugers M., Terzian Y., Chengalur J.~N., 1993, ApJ, 411, 778. doi:10.1086/172881

\bibitem[\protect\citeauthoryear{Bear \& Soker}{2025}]{BearSoker2025}  Bear E., Soker N., 2025, arXiv, arXiv:2407.03182. doi:10.48550/arXiv.2407.03182 

\bibitem[{{Berm{\'u}dez-Bustamante} {et~al.}(2024{\natexlab{a}}){Berm{\'u}dez-Bustamante}, {De Marco}, {Siess}, {Price}, {Gonz{\'a}lez-Bol{\'\i}var}, {Lau}, {Mu}, {Hirai}, {Danilovich}, \& {Kasliwal}}]{BermudezBustamanteetal2024b}
{Berm{\'u}dez-Bustamante}, L.~C., {De Marco}, O., {Siess}, L., {et~al.} 2024{\natexlab{a}}, \mnras, 533, 464, \dodoi{10.1093/mnras/stae1841}

\bibitem[{{Berm{\'u}dez-Bustamante} {et~al.}(2024{\natexlab{b}}){Berm{\'u}dez-Bustamante}, {De Marco}, {Siess}, {Price}, {Gonz{\'a}lez-Bol{\'\i}var}, {Lau}, {Mu}, {Hirai}, {Danilovich}, \& {Kasliwal}}]{BermudezBustamanteetal2024}
---. 2024{\natexlab{b}}, \mnras, 533, 464, \dodoi{10.1093/mnras/stae1841}

\bibitem[{{Bhattacharjee} {et~al.}(2024){Bhattacharjee}, {Kulkarni}, {Kong}, {Tam}, {Bond}, {El-Badry}, {Caiazzo}, {Graham}, {Rodriguez}, {Zeimann}, {Fremling}, {Drake}, {Werner}, {Rodriguez}, {Prince}, {Laher}, {Chen}, \& {Riddle}}]{Bhattacharjeeetal2024}
{Bhattacharjee}, S., {Kulkarni}, S.~R., {Kong}, A. K.~H., {et~al.} 2024, arXiv e-prints, arXiv:2410.03589, \dodoi{10.48550/arXiv.2410.03589}

\bibitem[{{Blackman} \& {Lucchini}(2014)}]{BlackmanLucchini2014}
{Blackman}, E.~G., \& {Lucchini}, S. 2014, \mnras, 440, L16, \dodoi{10.1093/mnrasl/slu001}

\bibitem[{{Boffin} \& {Jones}(2019)}]{BoffinJones2019}
{Boffin}, H. M.~J., \& {Jones}, D. 2019, {The Importance of Binaries in the Formation and Evolution of Planetary Nebulae}, \dodoi{10.1007/978-3-030-25059-1}

\bibitem[{{Boffin} {et~al.}(2012){Boffin}, {Miszalski}, {Rauch}, {Jones}, {Corradi}, {Napiwotzki}, {Day-Jones}, \& {K{\"o}ppen}}]{Boffinetal2012}
{Boffin}, H. M.~J., {Miszalski}, B., {Rauch}, T., {et~al.} 2012, Science, 338, 773, \dodoi{10.1126/science.1225386}

\bibitem[{{Brown} {et~al.}(2019){Brown}, {Jones}, {Boffin}, \& {Van Winckel}}]{Brownetal2019}
{Brown}, A.~J., {Jones}, D., {Boffin}, H. M.~J., \& {Van Winckel}, H. 2019, \mnras, 482, 4951, \dodoi{10.1093/mnras/sty2986}

\bibitem[{{Bujarrabal} {et~al.}(2001){Bujarrabal}, {Castro-Carrizo}, {Alcolea}, \& {S{\'a}nchez Contreras}}]{Bujarrabaletal2001}
{Bujarrabal}, V., {Castro-Carrizo}, A., {Alcolea}, J., \& {S{\'a}nchez Contreras}, C. 2001, \aap, 377, 868, \dodoi{10.1051/0004-6361:20011090}

\bibitem[{{Chamandy} {et~al.}(2020){Chamandy}, {Blackman}, {Frank}, {Carroll-Nellenback}, \& {Tu}}]{Chamandyetal2020}
{Chamandy}, L., {Blackman}, E.~G., {Frank}, A., {Carroll-Nellenback}, J., \& {Tu}, Y. 2020, \mnras, 495, 4028, \dodoi{10.1093/mnras/staa1273}

\bibitem[{{Chamandy} {et~al.}(2019){Chamandy}, {Blackman}, {Frank}, {Carroll-Nellenback}, {Zou}, \& {Tu}}]{Chamandyetal2019}
{Chamandy}, L., {Blackman}, E.~G., {Frank}, A., {et~al.} 2019, \mnras, 490, 3727, \dodoi{10.1093/mnras/stz2813}

\bibitem[{{Chamandy} {et~al.}(2024){Chamandy}, {Carroll-Nellenback}, {Blackman}, {Frank}, {Tu}, {Liu}, {Zou}, \& {Nordhaus}}]{Chamandyetal2024}
{Chamandy}, L., {Carroll-Nellenback}, J., {Blackman}, E.~G., {et~al.} 2024, \mnras, 528, 234, \dodoi{10.1093/mnras/stae036}

\bibitem[{{Chamandy} {et~al.}(2018){Chamandy}, {Frank}, {Blackman}, {Carroll-Nellenback}, {Liu}, {Tu}, {Nordhaus}, {Chen}, \& {Peng}}]{Chamandyetal2018}
{Chamandy}, L., {Frank}, A., {Blackman}, E.~G., {et~al.} 2018, \mnras, 480, 1898, \dodoi{10.1093/mnras/sty1950}

\bibitem[{{Clairmont} {et~al.}(2022){Clairmont}, {Steffen}, \& {Koning}}]{Clairmontetal2022}
{Clairmont}, R., {Steffen}, W., \& {Koning}, N. 2022, \mnras, 516, 2711, \dodoi{10.1093/mnras/stac2375}

\bibitem[{{Corradi} {et~al.}(2015){Corradi}, {Garc{\'\i}a-Rojas}, {Jones}, \& {Rodr{\'\i}guez-Gil}}]{Corradietal2015}
{Corradi}, R. L.~M., {Garc{\'\i}a-Rojas}, J., {Jones}, D., \& {Rodr{\'\i}guez-Gil}, P. 2015, \apj, 803, 99, \dodoi{10.1088/0004-637X/803/2/99}

\bibitem[{{Corradi} {et~al.}(1999){Corradi}, {Perinotto}, {Villaver}, {Mampaso}, \& {Gon{\c{c}}alves}}]{Corradietal1999}
{Corradi}, R. L.~M., {Perinotto}, M., {Villaver}, E., {Mampaso}, A., \& {Gon{\c{c}}alves}, D.~R. 1999, \apj, 523, 721, \dodoi{10.1086/307768}

\bibitem[{{Corradi} {et~al.}(2011){Corradi}, {Sabin}, {Miszalski}, {Rodr{\'\i}guez-Gil}, {Santander-Garc{\'\i}a}, {Jones}, {Drew}, {Mampaso}, {Barlow}, {Rubio-D{\'\i}ez}, {Casares}, {Viironen}, {Frew}, {Giammanco}, {Greimel}, \& {Sale}}]{Corradietal2011}
{Corradi}, R.~L.~M., {Sabin}, L., {Miszalski}, B., {et~al.} 2011, \mnras, 410, 1349, \dodoi{10.1111/j.1365-2966.2010.17523.x}

\bibitem[{{Corradi} {et~al.}(2014){Corradi}, {Rodr{\'\i}guez-Gil}, {Jones}, {Garc{\'\i}a-Rojas}, {Mampaso}, {Garc{\'\i}a-Alvarez}, {Pursimo}, {Eenm{\"a}e}, {Liimets}, \& {Miszalski}}]{Corradietal2014}
{Corradi}, R.~L.~M., {Rodr{\'\i}guez-Gil}, P., {Jones}, D., {et~al.} 2014, \mnras, 441, 2799, \dodoi{10.1093/mnras/stu703}

\bibitem[{{Danehkar}(2022)}]{Danehkar2022}
{Danehkar}, A. 2022, \apjs, 260, 14, \dodoi{10.3847/1538-4365/ac5cca}

\bibitem[\protect\citeauthoryear{De Marco et al.}{2011}]{DeMarcoetal2011} { De Marco O., Passy J.-C., Moe M., Herwig F., Mac Low M.-M., Paxton B., 2011, MNRAS, 411, 2277. doi:10.1111/j.1365-2966.2010.17891.x  }

\bibitem[{{Derlopa} {et~al.}(2024){Derlopa}, {Akras}, {Amram}, {Boumis}, {Chiotellis}, \& {de Oliveira}}]{Derlopaetal2024}
{Derlopa}, S., {Akras}, S., {Amram}, P., {et~al.} 2024, \mnras, 530, 3327, \dodoi{10.1093/mnras/stae1013}

\bibitem[{{Estrella-Trujillo} {et~al.}(2019){Estrella-Trujillo}, {Hern{\'a}ndez-Mart{\'\i}nez}, {Vel{\'a}zquez}, {Esquivel}, \& {Raga}}]{EstrellaTrujilloetal2019}
{Estrella-Trujillo}, D., {Hern{\'a}ndez-Mart{\'\i}nez}, L., {Vel{\'a}zquez}, P.~F., {Esquivel}, A., \& {Raga}, A.~C. 2019, \apj, 876, 29, \dodoi{10.3847/1538-4357/ab12e1}

\bibitem[{{Exter} {et~al.}(2003){Exter}, {Pollacco}, \& {Bell}}]{Exteretal2003}
{Exter}, K.~M., {Pollacco}, D.~L., \& {Bell}, S.~A. 2003, \mnras, 341, 1349, \dodoi{10.1046/j.1365-8711.2003.06505.x}

\bibitem[{{Gagnier} \& {Pejcha}(2024)}]{GagnierPejcha2024}
{Gagnier}, D., \& {Pejcha}, O. 2024, \aap, 683, A4, \dodoi{10.1051/0004-6361/202348383}

\bibitem[\protect\citeauthoryear{Gagnier \& Pejcha}{2024}]{GagnierPejcha2025} Gagnier D., Pejcha O., 2024, arXiv, arXiv:2412.04419


\bibitem[{{Garc{\'\i}a-D{\'\i}az} {et~al.}(2008){Garc{\'\i}a-D{\'\i}az}, {L{\'o}pez}, {Garc{\'\i}a-Segura}, {Richer}, \& {Steffen}}]{GarciaDiazetal2008}
{Garc{\'\i}a-D{\'\i}az}, M.~T., {L{\'o}pez}, J.~A., {Garc{\'\i}a-Segura}, G., {Richer}, M.~G., \& {Steffen}, W. 2008, \apj, 676, 402, \dodoi{10.1086/527468}

\bibitem[{{Garc{\'\i}a-Segura} {et~al.}(2020){Garc{\'\i}a-Segura}, {Taam}, \& {Ricker}}]{GarciaSeguraetal2020}
{Garc{\'\i}a-Segura}, G., {Taam}, R.~E., \& {Ricker}, P.~M. 2020, \apj, 893, 150, \dodoi{10.3847/1538-4357/ab8006}

\bibitem[{{Garc{\'\i}a-Segura} {et~al.}(2021){Garc{\'\i}a-Segura}, {Taam}, \& {Ricker}}]{GarciaSeguraetal2021}
---. 2021, \apj, 914, 111, \dodoi{10.3847/1538-4357/abfc4e}

\bibitem[{{Glanz} \& {Perets}(2018)}]{GlanzPerets2018}
{Glanz}, H., \& {Perets}, H.~B. 2018, \mnras, 478, L12, \dodoi{10.1093/mnrasl/sly065}

\bibitem[{{Glanz} \& {Perets}(2021{\natexlab{a}})}]{GlanzPerets2021a}
---. 2021{\natexlab{a}}, \mnras, 500, 1921, \dodoi{10.1093/mnras/staa3242}

\bibitem[{{Glanz} \& {Perets}(2021{\natexlab{b}})}]{GlanzPerets2021b}
---. 2021{\natexlab{b}}, \mnras, 507, 2659, \dodoi{10.1093/mnras/stab2291}

\bibitem[{{G{\'o}mez-Mu{\~n}oz} {et~al.}(2019){G{\'o}mez-Mu{\~n}oz}, {Manchado}, {Bianchi}, {Manteiga}, \& {V{\'a}zquez}}]{GomezMunozetal2019}
{G{\'o}mez-Mu{\~n}oz}, M.~A., {Manchado}, A., {Bianchi}, L., {Manteiga}, M., \& {V{\'a}zquez}, R. 2019, \apj, 885, 84, \dodoi{10.3847/1538-4357/ab3fa7}

\bibitem[{{Gonz{\'a}lez-Bol{\'\i}var} {et~al.}(2024{\natexlab{a}}){Gonz{\'a}lez-Bol{\'\i}var}, {De Marco}, {Berm{\'u}dez-Bustamante}, {Siess}, \& {Price}}]{BermudezBustamanteetal2024a}
{Gonz{\'a}lez-Bol{\'\i}var}, M., {De Marco}, O., {Berm{\'u}dez-Bustamante}, L.~C., {Siess}, L., \& {Price}, D.~J. 2024{\natexlab{a}}, \mnras, 527, 9145, \dodoi{10.1093/mnras/stad3748}

\bibitem[{{Gonz{\'a}lez-Bol{\'\i}var} {et~al.}(2024{\natexlab{b}}){Gonz{\'a}lez-Bol{\'\i}var}, {De Marco}, {Berm{\'u}dez-Bustamante}, {Siess}, \& {Price}}]{GonzalezBolivaretal2024}
---. 2024{\natexlab{b}}, \mnras, 527, 9145, \dodoi{10.1093/mnras/stad3748}

\bibitem[{{Gonz{\'a}lez-Bol{\'\i}var} {et~al.}(2022){Gonz{\'a}lez-Bol{\'\i}var}, {De Marco}, {Lau}, {Hirai}, \& {Price}}]{GonzalezBolivaretal2022}
{Gonz{\'a}lez-Bol{\'\i}var}, M., {De Marco}, O., {Lau}, M. Y.~M., {Hirai}, R., \& {Price}, D.~J. 2022, \mnras, 517, 3181, \dodoi{10.1093/mnras/stac2301}

\bibitem[{{Guerrero} {et~al.}(2021){Guerrero}, {Cazzoli}, {Rechy-Garc{\'\i}a}, {Ramos-Larios}, {Montoro-Molina}, {G{\'o}mez-Gonz{\'a}lez}, {Toal{\'a}}, \& {Fang}}]{Guerreroetal2021}
{Guerrero}, M.~A., {Cazzoli}, S., {Rechy-Garc{\'\i}a}, J.~S., {et~al.} 2021, \apj, 909, 44, \dodoi{10.3847/1538-4357/abe2aa}

\bibitem[{{Guerrero} \& {Miranda}(2012)}]{GuerreroMiranda2012}
{Guerrero}, M.~A., \& {Miranda}, L.~F. 2012, \aap, 539, A47, \dodoi{10.1051/0004-6361/201117923}

\bibitem[{{Guerrero} {et~al.}(2020){Guerrero}, {Suzett Rechy-Garc{\'\i}a}, \& {Ortiz}}]{Guerreroetal2020}
{Guerrero}, M.~A., {Suzett Rechy-Garc{\'\i}a}, J., \& {Ortiz}, R. 2020, \apj, 890, 50, \dodoi{10.3847/1538-4357/ab61fa}

\bibitem[{{Gurjar} {et~al.}(2024){Gurjar}, {Chamandy}, {Zou}, {Blackman}, {Liu}, \& {Nordhaus}}]{Gurjareta2024eas}
{Gurjar}, D., {Chamandy}, L.~R., {Zou}, A., {et~al.} 2024, in EAS2024, European Astronomical Society Annual Meeting, 438

\bibitem[{{Hillel} {et~al.}(2022){Hillel}, {Schreier}, \& {Soker}}]{Hilleletal2022}
{Hillel}, S., {Schreier}, R., \& {Soker}, N. 2022, \mnras, 514, 3212, \dodoi{10.1093/mnras/stac1341}

\bibitem[{{Hillel} {et~al.}(2023){Hillel}, {Schreier}, \& {Soker}}]{Hilleletal2023}
---. 2023, \apj, 955, 7, \dodoi{10.3847/1538-4357/acf19a}

\bibitem[\protect\citeauthoryear{Hillwig et al.}{2010}]{Hillwigetal2010} { Hillwig T.~C., Bond H.~E., Af{\c{s}}ar M., De Marco O., 2010, AJ, 140, 319. doi:10.1088/0004-6256/140/2/319 }

\bibitem[{{Hillwig} {et~al.}(2015){Hillwig}, {Frew}, {Louie}, {De Marco}, {Bond}, {Jones}, \& {Schaub}}]{Hillwigetal2015}
{Hillwig}, T.~C., {Frew}, D.~J., {Louie}, M., {et~al.} 2015, \aj, 150, 30, \dodoi{10.1088/0004-6256/150/1/30}

\bibitem[{{Hillwig} {et~al.}(2017){Hillwig}, {Frew}, {Reindl}, {Rotter}, {Webb}, \& {Margheim}}]{Hillwigetal2017}
{Hillwig}, T.~C., {Frew}, D.~J., {Reindl}, N., {et~al.} 2017, \aj, 153, 24, \dodoi{10.3847/1538-3881/153/1/24}

\bibitem[{{Hillwig} {et~al.}(2016){Hillwig}, {Jones}, {De Marco}, {Bond}, {Margheim}, \& {Frew}}]{Hillwigetal2016}
{Hillwig}, T.~C., {Jones}, D., {De Marco}, O., {et~al.} 2016, \apj, 832, 125, \dodoi{10.3847/0004-637X/832/2/125}

\bibitem[{{Hillwig} {et~al.}(2022){Hillwig}, {Reindl}, {Rotter}, {Rengstorf}, {Heber}, \& {Irrgang}}]{Hillwigetal2022H}
{Hillwig}, T.~C., {Reindl}, N., {Rotter}, H.~M., {et~al.} 2022, \mnras, 511, 2033, \dodoi{10.1093/mnras/stac226}

\bibitem[{{Huckvale} {et~al.}(2013){Huckvale}, {Prouse}, {Jones}, {Lloyd}, {Pollacco}, {L{\'o}pez}, {O'Brien}, {Sabin}, \& {Vaytet}}]{Huckvaleetal2013}
{Huckvale}, L., {Prouse}, B., {Jones}, D., {et~al.} 2013, \mnras, 434, 1505, \dodoi{10.1093/mnras/stt1109}

\bibitem[{{Iaconi} \& {De Marco}(2019)}]{IaconiDeMarco2019}
{Iaconi}, R., \& {De Marco}, O. 2019, \mnras, 490, 2550, \dodoi{10.1093/mnras/stz2756}

\bibitem[{{Iaconi} {et~al.}(2019){Iaconi}, {Maeda}, {De Marco}, {Nozawa}, \& {Reichardt}}]{Iaconietal2019}
{Iaconi}, R., {Maeda}, K., {De Marco}, O., {Nozawa}, T., \& {Reichardt}, T. 2019, \mnras, 489, 3334, \dodoi{10.1093/mnras/stz2312}

\bibitem[{{Iaconi} {et~al.}(2020){Iaconi}, {Maeda}, {Nozawa}, {De Marco}, \& {Reichardt}}]{Iaconietal2020}
{Iaconi}, R., {Maeda}, K., {Nozawa}, T., {De Marco}, O., \& {Reichardt}, T. 2020, \mnras, 497, 3166, \dodoi{10.1093/mnras/staa2169}

\bibitem[{{Iaconi} {et~al.}(2017){Iaconi}, {Reichardt}, {Staff}, {De Marco}, {Passy}, {Price}, {Wurster}, \& {Herwig}}]{Iaconietal2017b}
{Iaconi}, R., {Reichardt}, T., {Staff}, J., {et~al.} 2017, \mnras, 464, 4028, \dodoi{10.1093/mnras/stw2377}

\bibitem[{{Igoshev} {et~al.}(2020){Igoshev}, {Perets}, \& {Michaely}}]{Igoshevetal2020}
{Igoshev}, A.~P., {Perets}, H.~B., \& {Michaely}, E. 2020, \mnras, 494, 1448, \dodoi{10.1093/mnras/staa833}

\bibitem[{{Jones}(2020)}]{Jones2020Galax}
{Jones}, D. 2020, Galaxies, 8, 28, \dodoi{10.3390/galaxies8020028}

\bibitem[{{Jones}(2024)}]{Jones2025}
---. 2024, arXiv e-prints, arXiv:2411.06831, \dodoi{10.48550/arXiv.2411.06831}

\bibitem[{{Jones} \& {Boffin}(2017)}]{JonesBoffin2017}
{Jones}, D., \& {Boffin}, H. M.~J. 2017, Nature Astronomy, 1, 0117, \dodoi{10.1038/s41550-017-0117}

\bibitem[{{Jones} {et~al.}(2014){Jones}, {Boffin}, {Miszalski}, {Wesson}, {Corradi}, \& {Tyndall}}]{Jonesetal2014Hen}
{Jones}, D., {Boffin}, H.~M.~J., {Miszalski}, B., {et~al.} 2014, \aap, 562, A89, \dodoi{10.1051/0004-6361/201322797}

\bibitem[{{Jones} {et~al.}(2015){Jones}, {Boffin}, {Rodr{\'\i}guez-Gil}, {Wesson}, {Corradi}, {Miszalski}, \& {Mohamed}}]{Jonesetal2015}
{Jones}, D., {Boffin}, H.~M.~J., {Rodr{\'\i}guez-Gil}, P., {et~al.} 2015, \aap, 580, A19, \dodoi{10.1051/0004-6361/201425454}

\bibitem[{{Jones} {et~al.}(2019){Jones}, {Boffin}, {Sowicka}, {Miszalski}, {Rodr{\'\i}guez-Gil}, {Santander-Garc{\'\i}a}, \& {Corradi}}]{Jonesetal2019}
{Jones}, D., {Boffin}, H. M.~J., {Sowicka}, P., {et~al.} 2019, \mnras, 482, L75, \dodoi{10.1093/mnrasl/sly142}

\bibitem[{{Jones} {et~al.}(2023){Jones}, {Hillwig}, \& {Reindl}}]{Jonesetal2023}
{Jones}, D., {Hillwig}, T.~C., \& {Reindl}, N. 2023, in Highlights on Spanish Astrophysics XI, ed. M.~{Manteiga}, L.~{Bellot}, P.~{Benavidez}, A.~{de Lorenzo-C{\'a}ceres}, M.~A. {Fuente}, M.~J. {Mart{\'\i}nez}, M.~{V{\'a}zquez Acosta}, \& C.~{Dafonte}, 216, \dodoi{10.48550/arXiv.2304.06355}

\bibitem[{{Jones} {et~al.}(2016){Jones}, {Wesson}, {Garc{\'\i}a-Rojas}, {Corradi}, \& {Boffin}}]{Jonesetal2016}
{Jones}, D., {Wesson}, R., {Garc{\'\i}a-Rojas}, J., {Corradi}, R.~L.~M., \& {Boffin}, H.~M.~J. 2016, \mnras, 455, 3263, \dodoi{10.1093/mnras/stv2519}

\bibitem[{{Jones} {et~al.}(2010){Jones}, {Lloyd}, {Santander-Garc{\'\i}a}, {L{\'o}pez}, {Meaburn}, {Mitchell}, {O'Brien}, {Pollacco}, {Rubio-D{\'\i}ez}, \& {Vaytet}}]{Jonesetal2010}
{Jones}, D., {Lloyd}, M., {Santander-Garc{\'\i}a}, M., {et~al.} 2010, \mnras, 408, 2312, \dodoi{10.1111/j.1365-2966.2010.17277.x}

\bibitem[{{Jones} {et~al.}(2020){Jones}, {Boffin}, {Hibbert}, {Steinmetz}, {Wesson}, {Hillwig}, {Sowicka}, {Corradi}, {Garc{\'\i}a-Rojas}, {Rodr{\'\i}guez-Gil}, \& {Munday}}]{Jonesetal2020}
{Jones}, D., {Boffin}, H.~M.~J., {Hibbert}, J., {et~al.} 2020, \aap, 642, A108, \dodoi{10.1051/0004-6361/202038778}

\bibitem[{{Jones} {et~al.}(2022){Jones}, {Munday}, {Corradi}, {Rodr{\'\i}guez-Gil}, {Boffin}, {Zak}, {Sowicka}, {Parsons}, {Dhillon}, {Littlefair}, {Marsh}, {Reindl}, \& {Garc{\'\i}a-Rojas}}]{Jonesetal2022}
{Jones}, D., {Munday}, J., {Corradi}, R. L.~M., {et~al.} 2022, \mnras, 510, 3102, \dodoi{10.1093/mnras/stab3736}

\bibitem[{{Kimeswenger} {et~al.}(2021){Kimeswenger}, {Thorstensen}, {Fesen}, {Drechsler}, {Strottner}, {Germiniani}, {Steindl}, {Przybilla}, {Weil}, \& {Rupert}}]{Kimeswengeretal2021}
{Kimeswenger}, S., {Thorstensen}, J.~R., {Fesen}, R.~A., {et~al.} 2021, \aap, 656, A145, \dodoi{10.1051/0004-6361/202039787}

\bibitem[{{Kuruwita} {et~al.}(2016){Kuruwita}, {Staff}, \& {De Marco}}]{Kuruwitaetal2016}
{Kuruwita}, R.~L., {Staff}, J., \& {De Marco}, O. 2016, \mnras, 461, 486, \dodoi{10.1093/mnras/stw1414}

\bibitem[{{Landri} {et~al.}(2024){Landri}, {Ricker}, {Renzo}, {Rau}, \& {Vigna-G{\'o}mez}}]{Landrietal2024}
{Landri}, C., {Ricker}, P.~M., {Renzo}, M., {Rau}, S., \& {Vigna-G{\'o}mez}, A. 2024, arXiv e-prints, arXiv:2407.15932, \dodoi{10.48550/arXiv.2407.15932}

\bibitem[{{Lau} {et~al.}(2022{\natexlab{a}}){Lau}, {Hirai}, {Gonz{\'a}lez-Bol{\'\i}var}, {Price}, {De Marco}, \& {Mandel}}]{Lauetal2022a}
{Lau}, M. Y.~M., {Hirai}, R., {Gonz{\'a}lez-Bol{\'\i}var}, M., {et~al.} 2022{\natexlab{a}}, \mnras, 512, 5462, \dodoi{10.1093/mnras/stac049}

\bibitem[{{Lau} {et~al.}(2022{\natexlab{b}}){Lau}, {Hirai}, {Price}, \& {Mandel}}]{Lauetal2022b}
{Lau}, M. Y.~M., {Hirai}, R., {Price}, D.~J., \& {Mandel}, I. 2022{\natexlab{b}}, \mnras, 516, 4669, \dodoi{10.1093/mnras/stac2490}

\bibitem[{{Law-Smith} {et~al.}(2020){Law-Smith}, {Everson}, {Ramirez-Ruiz}, {de Mink}, {van Son}, {G{\"o}tberg}, {Zellmann}, {Vigna-G{\'o}mez}, {Renzo}, {Wu}, {Schr{\o}der}, {Foley}, \& {Hutchinson-Smith}}]{LawSmithetal2020}
{Law-Smith}, J. A.~P., {Everson}, R.~W., {Ramirez-Ruiz}, E., {et~al.} 2020, arXiv e-prints, arXiv:2011.06630, \dodoi{10.48550/arXiv.2011.06630}

\bibitem[{{Lopez} {et~al.}(1993){Lopez}, {Roth}, \& {Tapia}}]{Lopezetal1993}
{Lopez}, J.~A., {Roth}, M., \& {Tapia}, M. 1993, \aap, 267, 194

\bibitem[{{L{\'o}pez-C{\'a}mara} {et~al.}(2019){L{\'o}pez-C{\'a}mara}, {De Colle}, \& {Moreno M{\'e}ndez}}]{LopezCamaraetal2019}
{L{\'o}pez-C{\'a}mara}, D., {De Colle}, F., \& {Moreno M{\'e}ndez}, E. 2019, \mnras, 482, 3646, \dodoi{10.1093/mnras/sty2959}

\bibitem[{{L{\'o}pez-C{\'a}mara} {et~al.}(2022){L{\'o}pez-C{\'a}mara}, {De Colle}, {Moreno M{\'e}ndez}, {Shiber}, \& {Iaconi}}]{LopezCamaraetal2022}
{L{\'o}pez-C{\'a}mara}, D., {De Colle}, F., {Moreno M{\'e}ndez}, E., {Shiber}, S., \& {Iaconi}, R. 2022, \mnras, 513, 3634, \dodoi{10.1093/mnras/stac932}

\bibitem[{{L{\'o}pez-C{\'a}mara} {et~al.}(2020){L{\'o}pez-C{\'a}mara}, {Moreno M{\'e}ndez}, \& {De Colle}}]{LopezCamaraetal2020MN}
{L{\'o}pez-C{\'a}mara}, D., {Moreno M{\'e}ndez}, E., \& {De Colle}, F. 2020, \mnras, 497, 2057, \dodoi{10.1093/mnras/staa1983}

\bibitem[{{L{\"u}} {et~al.}(2013){L{\"u}}, {Zhu}, \& {Podsiadlowski}}]{Luetal2013}
{L{\"u}}, G., {Zhu}, C., \& {Podsiadlowski}, P. 2013, \apj, 768, 193, \dodoi{10.1088/0004-637X/768/2/193}

\bibitem[{{Manchado} {et~al.}(1996){Manchado}, {Guerrero}, {Stanghellini}, \& {Serra-Ricart}}]{Manchadoetal1996}
{Manchado}, A., {Guerrero}, M.~A., {Stanghellini}, L., \& {Serra-Ricart}, M. 1996, {The IAC morphological catalog of northern Galactic planetary nebulae}

\bibitem[\protect\citeauthoryear{Manchado, Stanghellini, \& Guerrero}{1996}]{Manchadoetal1996b} Manchado A., Stanghellini L., Guerrero M.~A., 1996, ApJL, 466, L95. doi:10.1086/310170

\bibitem[{{Meaburn} {et~al.}(2013){Meaburn}, {Boumis}, \& {Akras}}]{Meaburnetal2013}
{Meaburn}, J., {Boumis}, P., \& {Akras}, S. 2013, \mnras, 435, 3462, \dodoi{10.1093/mnras/stt1580}

\bibitem[{{Michaely} \& {Perets}(2019)}]{MichaelyPerets2019}
{Michaely}, E., \& {Perets}, H.~B. 2019, \mnras, 484, 4711, \dodoi{10.1093/mnras/stz352}

\bibitem[{{Miranda} {et~al.}(2024){Miranda}, {V{\'a}zquez}, {Olgu{\'\i}n}, {Guill{\'e}n}, \& {Mat{\'\i}as}}]{Mirandaetal2024}
{Miranda}, L.~F., {V{\'a}zquez}, R., {Olgu{\'\i}n}, L., {Guill{\'e}n}, P.~F., \& {Mat{\'\i}as}, J.~M. 2024, \aap, 687, A123, \dodoi{10.1051/0004-6361/202348173}

\bibitem[{{Miszalski} {et~al.}(2008){Miszalski}, {Acker}, {Moffat}, {Parker}, \& {Udalski}}]{Miszalskietal2008}
{Miszalski}, B., {Acker}, A., {Moffat}, A.~F.~J., {Parker}, Q.~A., \& {Udalski}, A. 2008, \aap, 488, L79, \dodoi{10.1051/0004-6361:200810529}

\bibitem[{{Miszalski} {et~al.}(2009){Miszalski}, {Acker}, {Parker}, \& {Moffat}}]{Miszalskietal2009II}
{Miszalski}, B., {Acker}, A., {Parker}, Q.~A., \& {Moffat}, A.~F.~J. 2009, \aap, 505, 249, \dodoi{10.1051/0004-6361/200912176}

\bibitem[{{Miszalski} {et~al.}(2013){Miszalski}, {Boffin}, \& {Corradi}}]{Miszalskietal2013}
{Miszalski}, B., {Boffin}, H. M.~J., \& {Corradi}, R. L.~M. 2013, \mnras, 428, L39, \dodoi{10.1093/mnrasl/sls011}

\bibitem[{{Miszalski} {et~al.}(2011{\natexlab{a}}){Miszalski}, {Corradi}, {Boffin}, {Jones}, {Sabin}, {Santander-Garc{\'\i}a}, {Rodr{\'\i}guez-Gil}, \& {Rubio-D{\'\i}ez}}]{Miszalskietal2011E}
{Miszalski}, B., {Corradi}, R.~L.~M., {Boffin}, H.~M.~J., {et~al.} 2011{\natexlab{a}}, \mnras, 413, 1264, \dodoi{10.1111/j.1365-2966.2011.18212.x}

\bibitem[{{Miszalski} {et~al.}(2011{\natexlab{b}}){Miszalski}, {Corradi}, {Jones}, {Santander-Garc{\'\i}a}, {Rodr{\'\i}guez-Gil}, \& {Rubio-D{\'\i}ez}}]{Miszalskietal2011apn5}
{Miszalski}, B., {Corradi}, R.~L.~M., {Jones}, D., {et~al.} 2011{\natexlab{b}}, in Asymmetric Planetary Nebulae 5 Conference, ed. A.~A. {Zijlstra}, F.~{Lykou}, I.~{McDonald}, \& E.~{Lagadec}, 328, \dodoi{10.48550/arXiv.1009.2890}

\bibitem[{{Miszalski} {et~al.}(2011{\natexlab{c}}){Miszalski}, {Jones}, {Rodr{\'\i}guez-Gil}, {Boffin}, {Corradi}, \& {Santander-Garc{\'\i}a}}]{Miszalskietal2011AA}
{Miszalski}, B., {Jones}, D., {Rodr{\'\i}guez-Gil}, P., {et~al.} 2011{\natexlab{c}}, \aap, 531, A158, \dodoi{10.1051/0004-6361/201117084}

\bibitem[{{Miszalski} {et~al.}(2018{\natexlab{a}}){Miszalski}, {Manick}, {Miko{\l}ajewska}, {I{\l}kiewicz}, {Kamath}, \& {Van Winckel}}]{Miszalskietal2018ngc1360}
{Miszalski}, B., {Manick}, R., {Miko{\l}ajewska}, J., {et~al.} 2018{\natexlab{a}}, \mnras, 473, 2275, \dodoi{10.1093/mnras/stx2501}

\bibitem[{{Miszalski} {et~al.}(2018{\natexlab{b}}){Miszalski}, {Manick}, {Miko{\l}ajewska}, {Van Winckel}, \& {I{\l}kiewicz}}]{Miszalski2018mycn18}
{Miszalski}, B., {Manick}, R., {Miko{\l}ajewska}, J., {Van Winckel}, H., \& {I{\l}kiewicz}, K. 2018{\natexlab{b}}, \pasa, 35, e027, \dodoi{10.1017/pasa.2018.23}

\bibitem[{{Miszalski} {et~al.}(2019{\natexlab{a}}){Miszalski}, {Manick}, {Rauch}, {I{\l}kiewicz}, {Van Winckel}, \& {Miko{\l}ajewska}}]{Miszalskietal2019sp}
{Miszalski}, B., {Manick}, R., {Rauch}, T., {et~al.} 2019{\natexlab{a}}, \pasa, 36, e042, \dodoi{10.1017/pasa.2019.36}

\bibitem[{{Miszalski} {et~al.}(2019{\natexlab{b}}){Miszalski}, {Manick}, {Van Winckel}, \& {Escorza}}]{Miszalskietal2019ngc2392}
{Miszalski}, B., {Manick}, R., {Van Winckel}, H., \& {Escorza}, A. 2019{\natexlab{b}}, \pasa, 36, e018, \dodoi{10.1017/pasa.2019.11}

\bibitem[{{Miszalski} {et~al.}(2019{\natexlab{c}}){Miszalski}, {Manick}, {Van Winckel}, \& {Miko{\l}ajewska}}]{Miszalski2019ic}
{Miszalski}, B., {Manick}, R., {Van Winckel}, H., \& {Miko{\l}ajewska}, J. 2019{\natexlab{c}}, \mnras, 487, 1040, \dodoi{10.1093/mnras/stz1315}

\bibitem[{{Mitchell} {et~al.}(2007){Mitchell}, {Pollacco}, {O'Brien}, {Bryce}, {L{\'o}pez}, {Meaburn}, \& {Vaytet}}]{Mitchelletal2007}
{Mitchell}, D.~L., {Pollacco}, D., {O'Brien}, T.~J., {et~al.} 2007, \mnras, 374, 1404, \dodoi{10.1111/j.1365-2966.2006.11251.x}

\bibitem[{{Moraga Baez} {et~al.}(2023){Moraga Baez}, {Kastner}, {Balick}, {Montez}, \& {Bublitz}}]{MoragaBaezetal2023}
{Moraga Baez}, P., {Kastner}, J.~H., {Balick}, B., {Montez}, R., \& {Bublitz}, J. 2023, \apj, 942, 15, \dodoi{10.3847/1538-4357/aca401}

\bibitem[{{Moreno} {et~al.}(2022){Moreno}, {Schneider}, {R{\"o}pke}, {Ohlmann}, {Pakmor}, {Podsiadlowski}, \& {Sand}}]{Morenoetal2022}
{Moreno}, M.~M., {Schneider}, F. R.~N., {R{\"o}pke}, F.~K., {et~al.} 2022, \aap, 667, A72, \dodoi{10.1051/0004-6361/202142731}

\bibitem[{{Moreno M{\'e}ndez} {et~al.}(2017){Moreno M{\'e}ndez}, {L{\'o}pez-C{\'a}mara}, \& {De Colle}}]{MorenoMendezetal2017}
{Moreno M{\'e}ndez}, E., {L{\'o}pez-C{\'a}mara}, D., \& {De Colle}, F. 2017, \mnras, 470, 2929, \dodoi{10.1093/mnras/stx1385}

\bibitem[{{Morris}(1987)}]{Morris1987}
{Morris}, M. 1987, \pasp, 99, 1115, \dodoi{10.1086/132089}

\bibitem[{{Nandez} {et~al.}(2014){Nandez}, {Ivanova}, \& {Lombardi}}]{Nandezetal2014}
{Nandez}, J.~L.~A., {Ivanova}, N., \& {Lombardi}, J.~C., J. 2014, \apj, 786, 39, \dodoi{10.1088/0004-637X/786/1/39}

\bibitem[{{Nordhaus} {et~al.}(2007){Nordhaus}, {Blackman}, \& {Frank}}]{Nordhausetal2007}
{Nordhaus}, J., {Blackman}, E.~G., \& {Frank}, A. 2007, \mnras, 376, 599, \dodoi{10.1111/j.1365-2966.2007.11417.x}

\bibitem[{{Noughani} {et~al.}(2024){Noughani}, {Nordhaus}, {Richmond}, \& {Wilson}}]{Noughanietal2025}
{Noughani}, N., {Nordhaus}, J., {Richmond}, M., \& {Wilson}, E.~C. 2024, arXiv e-prints, arXiv:2406.04118, \dodoi{10.48550/arXiv.2406.04118}

\bibitem[{{O'Connor} {et~al.}(2000){O'Connor}, {Redman}, {Holloway}, {Bryce}, {L{\'o}pez}, \& {Meaburn}}]{OConnoretal2000}
{O'Connor}, J.~A., {Redman}, M.~P., {Holloway}, A.~J., {et~al.} 2000, \apj, 531, 336, \dodoi{10.1086/308452}

\bibitem[{{Ohlmann} {et~al.}(2016){Ohlmann}, {R{\"o}pke}, {Pakmor}, \& {Springel}}]{Ohlmannetal2016a}
{Ohlmann}, S.~T., {R{\"o}pke}, F.~K., {Pakmor}, R., \& {Springel}, V. 2016, \apjl, 816, L9, \dodoi{10.3847/2041-8205/816/1/L9}

\bibitem[{{Ondratschek} {et~al.}(2022){Ondratschek}, {R{\"o}pke}, {Schneider}, {Fendt}, {Sand}, {Ohlmann}, {Pakmor}, \& {Springel}}]{Ondratscheketal2022}
{Ondratschek}, P.~A., {R{\"o}pke}, F.~K., {Schneider}, F. R.~N., {et~al.} 2022, \aap, 660, L8, \dodoi{10.1051/0004-6361/202142478}

\bibitem[{{Orosz} {et~al.}(2017){Orosz}, {Imai}, {Dodson}, {Rioja}, {Frey}, {Burns}, {Etoka}, {Nakagawa}, {Nakanishi}, {Asaki}, {Goldman}, \& {Tafoya}}]{Oroszetal2019}
{Orosz}, G., {Imai}, H., {Dodson}, R., {et~al.} 2017, \aj, 153, 119, \dodoi{10.3847/1538-3881/153/3/119}

\bibitem[{{Rechy-Garc{\'\i}a} {et~al.}(2020){Rechy-Garc{\'\i}a}, {Guerrero}, {Duarte Puertas}, {Chu}, {Toal{\'a}}, \& {Miranda}}]{RechyGarciaetal2020}
{Rechy-Garc{\'\i}a}, J.~S., {Guerrero}, M.~A., {Duarte Puertas}, S., {et~al.} 2020, \mnras, 492, 1957, \dodoi{10.1093/mnras/stz3326}

\bibitem[{{Reichardt} {et~al.}(2020){Reichardt}, {De Marco}, {Iaconi}, {Chamandy}, \& {Price}}]{Reichardtetal2020}
{Reichardt}, T.~A., {De Marco}, O., {Iaconi}, R., {Chamandy}, L., \& {Price}, D.~J. 2020, \mnras, 494, 5333, \dodoi{10.1093/mnras/staa937}

\bibitem[{{Rosselli-Calderon} {et~al.}(2024){Rosselli-Calderon}, {Yarza}, {Murguia-Berthier}, {Rohoza}, {Everson}, {Antoni}, {MacLeod}, \& {Ramirez-Ruiz}}]{RosselliCalderon2024}
{Rosselli-Calderon}, A., {Yarza}, R., {Murguia-Berthier}, A., {et~al.} 2024, arXiv e-prints, arXiv:2404.08037, \dodoi{10.48550/arXiv.2404.08037}

\bibitem[{{Sabach} {et~al.}(2017){Sabach}, {Hillel}, {Schreier}, \& {Soker}}]{Sabachetal2017}
{Sabach}, E., {Hillel}, S., {Schreier}, R., \& {Soker}, N. 2017, \mnras, 472, 4361, \dodoi{10.1093/mnras/stx2272}

\bibitem[{{Sahai} {et~al.}(2008){Sahai}, {Claussen}, {S{\'a}nchez Contreras}, {Morris}, \& {Sarkar}}]{Sahaietal2008}
{Sahai}, R., {Claussen}, M., {S{\'a}nchez Contreras}, C., {Morris}, M., \& {Sarkar}, G. 2008, \apj, 680, 483, \dodoi{10.1086/587638}

\bibitem[{{Sahai} {et~al.}(2011){Sahai}, {Morris}, \& {Villar}}]{Sahaietal2011}
{Sahai}, R., {Morris}, M.~R., \& {Villar}, G.~G. 2011, \aj, 141, 134, \dodoi{10.1088/0004-6256/141/4/134}

\bibitem[{{Sahai} \& {Trauger}(1998)}]{SahaiTrauger1998}
{Sahai}, R., \& {Trauger}, J.~T. 1998, \aj, 116, 1357, \dodoi{10.1086/300504}

\bibitem[{{Sahai} {et~al.}(2017){Sahai}, {Vlemmings}, \& {Nyman}}]{Sahaietal2017}
{Sahai}, R., {Vlemmings}, W.~H.~T., \& {Nyman}, L.~{\r{A}}. 2017, \apj, 841, 110, \dodoi{10.3847/1538-4357/aa6d86}

\bibitem[{{Sahai} {et~al.}(2024){Sahai}, {Alcolea}, {Balick}, {Blackman}, {Bujarrabal}, {Castro-Carrizo}, {De Marco}, {Kastner}, {Kim}, {Lagadec}, {Lee}, {Sabin}, {Santander-Garcia}, {S{\'a}nchez Contreras}, {Tafoya}, {Ueta}, {Vlemmings}, \& {Zijlstra}}]{Sahaietal2024}
{Sahai}, R., {Alcolea}, J., {Balick}, B., {et~al.} 2024, arXiv e-prints, arXiv:2409.06038.
\newblock \doarXiv{2409.06038}

\bibitem[{{Santander-Garc{\'\i}a} {et~al.}(2015){Santander-Garc{\'\i}a}, {Rodr{\'\i}guez-Gil}, {Corradi}, {Jones}, {Miszalski}, {Boffin}, {Rubio-D{\'\i}ez}, \& {Kotze}}]{SantanderGarciaetal2015}
{Santander-Garc{\'\i}a}, M., {Rodr{\'\i}guez-Gil}, P., {Corradi}, R.~L.~M., {et~al.} 2015, \nat, 519, 63, \dodoi{10.1038/nature14124}

\bibitem[{{Schreier} {et~al.}(2023){Schreier}, {Hillel}, \& {Soker}}]{Schreieretal2023}
{Schreier}, R., {Hillel}, S., \& {Soker}, N. 2023, \mnras, 520, 4182, \dodoi{10.1093/mnras/stad360}

\bibitem[{{Schwarz} {et~al.}(1992){Schwarz}, {Corradi}, \& {Melnick}}]{Schwarzetal199}
{Schwarz}, H.~E., {Corradi}, R.~L.~M., \& {Melnick}, J. 1992, \aaps, 96, 23

\bibitem[\protect\citeauthoryear{Scolnic et al.}{2025}]{Scolnicetal2025} {Scolnic, A., Bear E., Soker N., 2025, arXiv, to be posted on January 7 }

\bibitem[\protect\citeauthoryear{Shiber}{2018}]{Shiber2018} Shiber S., 2018, Galax, 6, 96. doi:10.3390/galaxies6030096

\bibitem[{{Shiber} \& {Iaconi}(2024)}]{ShiberIaconi2024}
{Shiber}, S., \& {Iaconi}, R. 2024, \mnras, 532, 692, \dodoi{10.1093/mnras/stae1500}

\bibitem[{{Shiber} {et~al.}(2019){Shiber}, {Iaconi}, {De Marco}, \& {Soker}}]{Shiberetal2019}
{Shiber}, S., {Iaconi}, R., {De Marco}, O., \& {Soker}, N. 2019, \mnras, 488, 5615, \dodoi{10.1093/mnras/stz2013}

\bibitem[{{Shiber} {et~al.}(2016){Shiber}, {Schreier}, \& {Soker}}]{Shiberetal2016}
{Shiber}, S., {Schreier}, R., \& {Soker}, N. 2016, Research in Astronomy and Astrophysics, 16, 117, \dodoi{10.1088/1674-4527/16/7/117}

\bibitem[{{Shiber} \& {Soker}(2018)}]{ShiberSoker2018}
{Shiber}, S., \& {Soker}, N. 2018, \mnras, 477, 2584, \dodoi{10.1093/mnras/sty843}

\bibitem[{{Soker}(1990)}]{Soker1990AJ}
{Soker}, N. 1990, \aj, 99, 1869, \dodoi{10.1086/115465}

\bibitem[{{Soker}(1992)}]{Soker1992funnel}
---. 1992, \apj, 389, 628, \dodoi{10.1086/171235}

\bibitem[{{Soker}(1998)}]{Soker1998dust}
---. 1998, \mnras, 299, 1242, \dodoi{10.1046/j.1365-8711.1998.01884.x}

\bibitem[{{Soker}(2000)}]{Soker2000dust}
---. 2000, \apj, 540, 436, \dodoi{10.1086/309326}

\bibitem[{{Soker}(2020)}]{Soker2020Galax}
---. 2020, Galaxies, 8, 26, \dodoi{10.3390/galaxies8010026}

\bibitem[{{Soker}(2022)}]{Soker2022Rev}
---. 2022, Research in Astronomy and Astrophysics, 22, 122003, \dodoi{10.1088/1674-4527/ac9782}

\bibitem[{{Sowicka} {et~al.}(2017){Sowicka}, {Jones}, {Corradi}, {Wesson}, {Garc{\'\i}a-Rojas}, {Santander-Garc{\'\i}a}, {Boffin}, \& {Rodr{\'\i}guez-Gil}}]{Sowickaetal2017}
{Sowicka}, P., {Jones}, D., {Corradi}, R. L.~M., {et~al.} 2017, \mnras, 471, 3529, \dodoi{10.1093/mnras/stx1697}

\bibitem[{{Staff} {et~al.}(2016){Staff}, {De Marco}, {Macdonald}, {Galaviz}, {Passy}, {Iaconi}, \& {Low}}]{Staffetal2016MN}
{Staff}, J.~E., {De Marco}, O., {Macdonald}, D., {et~al.} 2016, \mnras, 455, 3511, \dodoi{10.1093/mnras/stv2548}

\bibitem[{{Tafoya} {et~al.}(2019){Tafoya}, {Orosz}, {Vlemmings}, {Sahai}, \& {P{\'e}rez-S{\'a}nchez}}]{Tafoyaetal2019}
{Tafoya}, D., {Orosz}, G., {Vlemmings}, W.~H.~T., {Sahai}, R., \& {P{\'e}rez-S{\'a}nchez}, A.~F. 2019, \aap, 629, A8, \dodoi{10.1051/0004-6361/201834632}

\bibitem[{{Toal{\'a}} {et~al.}(2020){Toal{\'a}}, {Guerrero}, {Bianchi}, {Chu}, \& {De Marco}}]{Toalaetal2020}
{Toal{\'a}}, J.~A., {Guerrero}, M.~A., {Bianchi}, L., {Chu}, Y.~H., \& {De Marco}, O. 2020, \mnras, 494, 3784, \dodoi{10.1093/mnras/staa1024}

\bibitem[{{Tocknell} {et~al.}(2014){Tocknell}, {De Marco}, \& {Wardle}}]{Tocknelletal2014}
{Tocknell}, J., {De Marco}, O., \& {Wardle}, M. 2014, \mnras, 439, 2014, \dodoi{10.1093/mnras/stu079}

\bibitem[\protect\citeauthoryear{Van Winckel et al.}{2014}]{vanWinckeletal2014} { Van Winckel H., Jorissen A., Exter K., Raskin G., Prins S., Perez Padilla J., Merges F., et al., 2014, A\&A, 563, L10. }  

\bibitem[{{V{\'a}zquez} {et~al.}(2002){V{\'a}zquez}, {Miranda}, {Torrelles}, {Olgu{\'\i}n}, {Ben{\'\i}tez}, {Rodr{\'\i}guez}, \& {L{\'o}pez}}]{Vazquezetal2002}
{V{\'a}zquez}, R., {Miranda}, L.~F., {Torrelles}, J.~M., {et~al.} 2002, \apj, 576, 860, \dodoi{10.1086/341792}

\bibitem[{{Vetter} {et~al.}(2024){Vetter}, {Roepke}, {Schneider}, {Pakmor}, {Ohlmann}, {Lau}, \& {Andrassy}}]{Vetteretal2024}
{Vetter}, M., {Roepke}, F.~K., {Schneider}, F.~R.~N., {et~al.} 2024, arXiv e-prints, arXiv:2410.07841, \dodoi{10.48550/arXiv.2410.07841}

\bibitem[{{Wesson} {et~al.}(2018){Wesson}, {Jones}, {Garc{\'\i}a-Rojas}, {Boffin}, \& {Corradi}}]{Wessonetal2018}
{Wesson}, R., {Jones}, D., {Garc{\'\i}a-Rojas}, J., {Boffin}, H.~M.~J., \& {Corradi}, R.~L.~M. 2018, \mnras, 480, 4589, \dodoi{10.1093/mnras/sty1871}

\bibitem[{{Wilson} \& {Nordhaus}(2019)}]{WilsonNordhaus2019}
{Wilson}, E.~C., \& {Nordhaus}, J. 2019, \mnras, 485, 4492, \dodoi{10.1093/mnras/stz601}

\bibitem[{{Wilson} \& {Nordhaus}(2020)}]{WilsonNordhaus2020}
---. 2020, \mnras, 497, 1895, \dodoi{10.1093/mnras/staa2088}

\bibitem[{{Wilson} \& {Nordhaus}(2022)}]{WilsonNordhaus2022}
---. 2022, \mnras, 516, 2189, \dodoi{10.1093/mnras/stac2300}

\bibitem[{{Zou} {et~al.}(2022){Zou}, {Chamandy}, {Carroll-Nellenback}, {Blackman}, \& {Frank}}]{Zouetal2022}
{Zou}, Y., {Chamandy}, L., {Carroll-Nellenback}, J., {Blackman}, E.~G., \& {Frank}, A. 2022, \mnras, 514, 3041, \dodoi{10.1093/mnras/stac1529}

\bibitem[{{Zou} {et~al.}(2020){Zou}, {Frank}, {Chen}, {Reichardt}, {De Marco}, {Blackman}, {Nordhaus}, {Balick}, {Carroll-Nellenback}, {Chamandy}, \& {Liu}}]{Zouetal2020}
{Zou}, Y., {Frank}, A., {Chen}, Z., {et~al.} 2020, \mnras, 497, 2855, \dodoi{10.1093/mnras/staa2145}

\end{thebibliography}
\end{document}